\documentclass[12pt,a4paper]{article}
\usepackage{graphicx}
\usepackage{times}
\title{\bf  The nebulae around LBVs: \\ a multiwavelength approach }
\author{Grazia Umana$^1$, Carla S. Buemi$^1$, Corrado Trigilio$^1$, Paolo Leto$^1$ \\ Joseph  L. Hora$^2$ and Giovanni Fazio$^2$\\
\vspace{1cm}\\
\normalsize $^1$ INAF-Osservatorio Astrofisico di Catania, Via S. Sofia 78, 95123 Catania, ITALY\\ 
\normalsize $^2$ Harvard-Smithsonian Center for Astrophysics, 60 Garden St. MS-65, \\ \normalsize Cambridge, MA 02138-1516, USA \\}
\date{\mbox{}}
\begin{document}
\maketitle
\pagestyle{empty}
%
%
\def\bull{\vrule height .9ex width .8ex depth -.1ex}
\makeatletter
\def\ps@plain{\let\@mkboth\gobbletwo
\def\@oddhead{}\def\@oddfoot{\hfil\tiny\bull\quad
``The multi-wavelength view of hot, massive stars''; 39$^{\rm th}$ Li\`ege Int.\ Astroph.\ Coll., 12-16 July 2010 \quad\bull}%
\def\@evenhead{}\let\@evenfoot\@oddfoot}
\makeatother
%
%
\def\beginrefer{\section*{References}%
\begin{quotation}\mbox{}\par}
\def\refer#1\par{{\setlength{\parindent}{-\leftmargin}\indent#1\par}}
\def\endrefer{\end{quotation}}
%
%
{\noindent\small{\bf Abstract:} 
We present first results of our study of a sample of Galactic LBV, 
aimed to contribute to a better understanding of the LBV phenomenon, 
by recovering  the mass-loss history of the central object from the analysis 
of its associated nebula.
Mass-loss properties have been derived  by a synergistic use of 
different techniques, at different wavelengths, to
obtain high-resolution, multi-wavelength maps, tracing the 
different emitting components coexisting in the stellar ejecta:
the ionized/neutral gas and the dust. 
Evidence for asymmetric mass-loss and observational evidence of
possible mutual interaction between gas and dust components have 
been observed by the comparison of  mid-IR (Spitzer/IRAC , VLT/VISIR) 
and radio (VLA) images of the nebulae, while
important information on the gas and dust composition 
have been derived from Spitzer/IRS spectra.

}
%
%
\section{Introduction}
Luminous Blue Variables are luminous (intrinsically bright, $ L \sim 10^{4} L_{\odot}$),
which show different kinds of photometric and spectroscopic variabilities. They are massive
($M \sim 22-120  M_{\odot}$), characterized by  intense mass-loss rates ($10^{-6}-10^{-4} M_{\odot} yr^{-1}$), which can occur also in the form of eruptive events. 
LBVs are quite rare objects in our Galaxy. This is probably connected to their very short lifetime (some $10^{4} yrs$).
The most recent census of Galactic LBVs counts 12 effective members and 23 candidates (Clark et al. 2005) and  a
few LBV (and candidates) have been also reported in some nearby galaxies.

LBVs represent a crucial phase in massive star evolution during which a star loses enough mass to become a $\sim 20  M_{\odot}$ WR star.  To test evolutionary models, it is extremely important to quantify a key parameter: the total mass lost during the LBV phase, i.e. the  gas (ionized, neutral, and molecular, if it exists) and the dust.
Another important aspect of the study of circumstellar envelopes is to determine the mass-loss archeology of central star and in particular  how  the mass-loss behavior (multiple events, bursts) is related to the physical parameters of the central object.

\section{The project}
A good  understanding of the physical conditions in LBV ejecta requires multi-wavelength observations, tracing the 
different emitting components coexisting in the stellar ejecta:
the ionized/neutral gas and the dust. The study of both components provides two kinds of information: current mass-loss, via direct observations of 
stellar winds (the gas component), and mass loss history of the central star, by analysis of the dust component/s.
The detailed knowledge of the gas and dust distribution  allows us to evaluate the total (gas+dust) mass of the nebula, 
the presence of different shells related to different mass-loss episodes, and thus  
the total mass lost by the central object during this critical phase of its evolution. 
Moreover, it could provide evidence for gas and dust mutual interactions which are a possible cause of the quite complex morphologies often observed in the LBVNs.

In the last few years we have started a systematic study of a sample of Galactic LBVs and LBV candidates aimed at deriving their   mass-loss properties for a better understanding of the LBV phenomenon in the wider context of massive star evolution. 
Our approach is based on a synergistic use of different techniques, at different wavelengths, that allows us to analyze the several emitting components coexisting in the nebula. In particular, we performed a detailed comparison of mid-IR and radio  maps, with comparable spatial resolution, 
to sort out the spatial differences in the maps in order to detect particular features which can be associated 
with mass-loss during the LBV phase: asymmetric winds versus symmetric winds in asymmetric environments; single events versus multiple events.
In the framework of our LBV project, we obtained Spitzer/IRAC  observations aimed at detecting and resolving the faint dust shells ejected from the central
stars,  and Spitzer/IRS observations to characterize the dust content of the nebula via
mid-IR spectra.  The ionised fraction of the nebulae has been mapped  via high-angular resolution radio  (VLA) observations.
For the more compact nebulae, images in the mid-IR have been obtained by using  VLT/VISIR in the N and Q  mid-IR bands.

\section{Results}

Many interesting results have been obtained from our imaging and spectroscopic program.
In particular,  extended dusty shells have been detected around some of our targets
(see Fig. 1) and evidence for asymmetric mass-loss and of
possible mutual interaction between gas and dust components is suggested  
by the comparison of  VLT/VISIR and VLA images of the more compact nebulae.
The analysis of  the mid-IR spectra has provided  information on the gas and dust composition,
allowing  identification of the mineral composition of 
LBV ejecta and to discriminate between crystalline or amorphous dust components.
Moreover,  the presence  of low-excitation atomic fine structure lines points out the existence  of a   photodissociation region (PDR),
 an extra component of neutral/molecular gas  that should be taken into account when one determines 
 the total budget of mass lost by the star during its LBV phase.
We present examples of our results in the following sections. More details  can be found in a series of papers devoted to the project.
\begin{figure}
\centering
\includegraphics[scale=0.25]{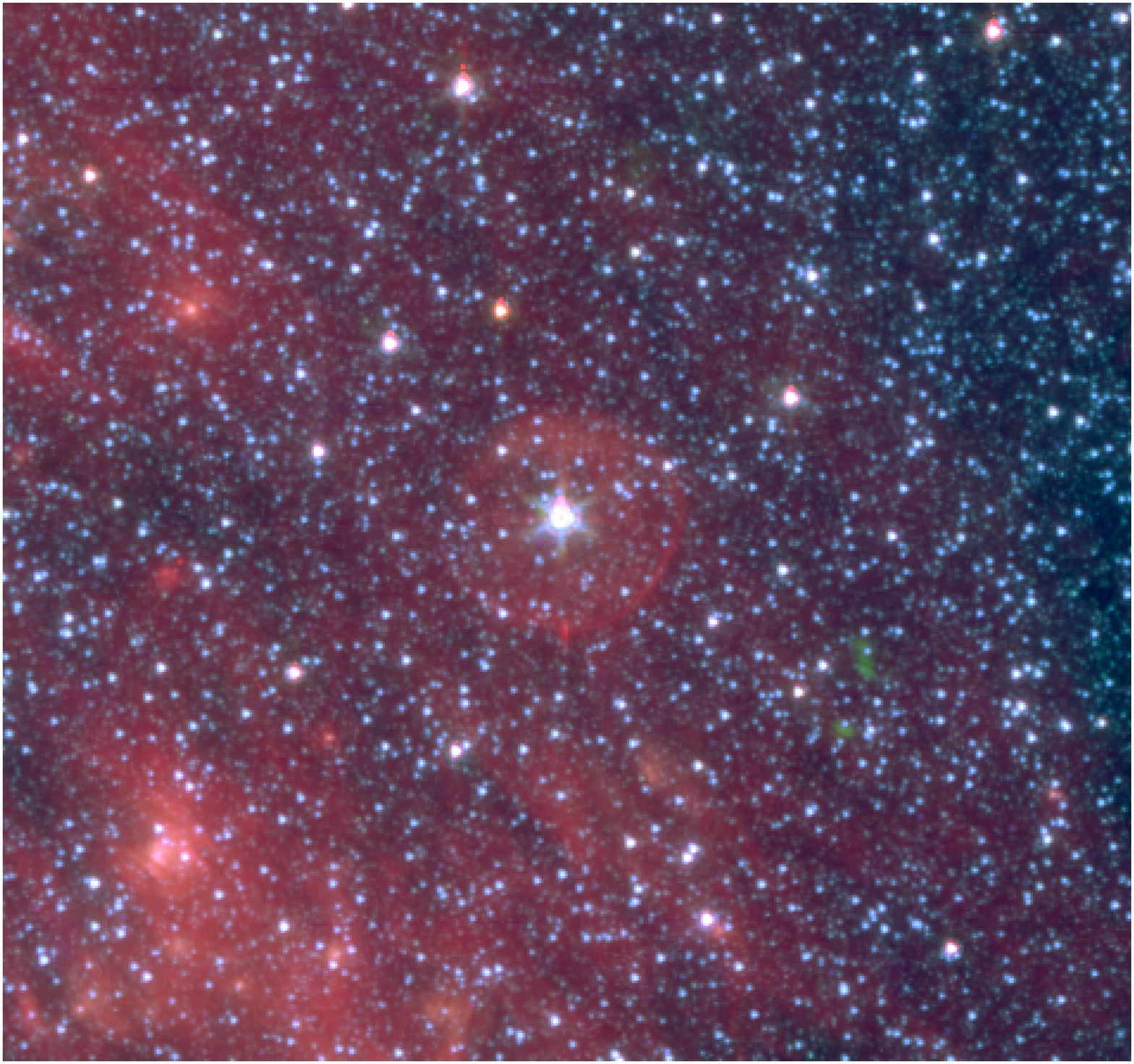}
\caption{IRAC composite image of Wray 17-96 with north up and east to the left, $FOV=5^{\prime \prime}$.
Emission from the central star is evident at $3.6 \mu m$ (Blue), while the warm dust is well traced by the 
 $8.5 \mu m$  (Red)}
\end{figure}

\subsection{IRAS 18576+034}
This is the object which shows the most extreme difference between the ionized gas component
(traced by free-free emission ) and the dust component (Fig. 2). 
High spatial resolution and high sensitivity images of  IRAS 18576+0341
were obtained using the mid infrared imager VISIR at the Very Large Telescope and the  Very Large Array interferometer (see Buemi et al., 2010 for details)
The approximately circularly-symmetric, mid-IR nebula strongly contrasts  with the asymmetry that characterizes the ionized component of the envelope, 
as seen in the radio and [Ne\,{\sc ii}] line images.
Among possible scenarios for the cause of the observed asymmetry in the ionized gas morphology are 
an unseen external ionizing source (either a companion or shocked gas)  or  holes in the dusty material. However,  at the moment  it is not possible to discriminate amongst them. 

The detailed mid-IR maps allowed us to determine the  size of the dusty nebula
($ R_{dust}$=$7^{\prime \prime}$), the dust temperature distribution, and the total dust mass. 
From the total dust mass (Buemi et al. 2010),  assuming a gas to dust ratio of 100, a total nebular mass of $\sim 0.5 M_{\odot}$ is derived. If this material is expanding at 
$70\ km/sec$ (Clark et al. 2009), we derive an averaged mass-loss rate 
of $\sim 1 \times 10^{-4}   M\odot$yr$^{-1}$ during the nebula formation.
However, the dust distribution in the nebula is consistent with a strong mass-loss episode that occurred $\sim$ 2000 years ago (Buemi et al. 2010),  indicating the mass-loss is not constant, with different quantity of mass released during  episodes of different duration.
This result is corroborated by the current-day mass loss rate of  $3.7  \times 10^{-5} 
M\odot$yr$^{-1}$  from the central object as measured in the radio (Umana et al. 2005), which is smaller than the average value necessary to fill up the circumstellar nebula.

\begin{figure}
\centering
\includegraphics[scale=0.7]{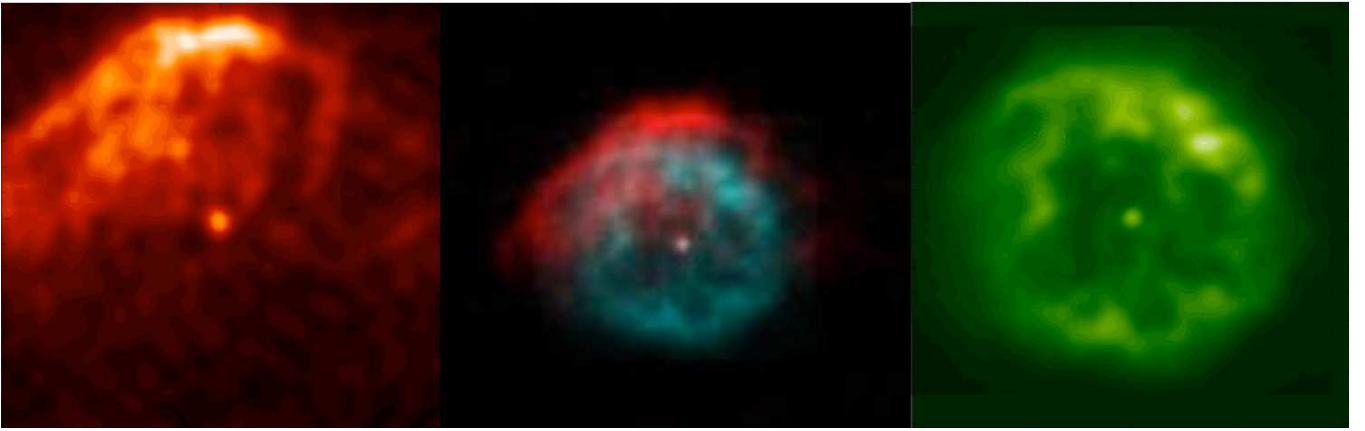}
\caption{(center) Multi-configuration 6-$cm$  VLA map  (red) superimposed on the 11.26-$ \mu m$ VISIR map (blue) of IRAS~18576+0341.
Both maps have north up and east to the left.
The center image is $20^{\prime \prime}$ across. A zoomed (FOV  $8^{\prime \prime}$ ) of the VLA map (left) and of the VISIR map (right) are also shown (adapted from Buemi et al. 2010). }
\end{figure}

\subsection{HR Car}
HR Car is surrounded by a faint, low-excitation nebula which is difficult to observe because of the high luminosity of the central object. One of the most striking properties of the nebula is the difference between the large scale optical  morphology and the inner, strongly asymmetric,  radio nebula (see White 2000).
Our spectroscopic  Spitzer/IRS observations
of the inner nebula reveal a rich mid-IR spectrum showing both solid state and atomic gas signatures (Umana et al. 2009).
The characteristic broad feature at $10\mu m$  indicates the presence of amorphous silicates, suggesting that dust formation occurred  during the LBV outburst.  
This is in contrast with the detection of crystalline dust in other 
Galactic LBVs that are probably more evolved. The crystalline dust is similar to the dust observed in red
supergiants  that  has been considered to be evidence of dust production
during evolutionary phases prior to the outburst  (Waters et al. 1998).
Strong low-excitation atomic fine structure lines such as $ 26.0\mu m$
[Fe\,{\sc ii}]  and $ 34.8\mu m$ [Si\,{\sc ii}]  indicate, for the first
time, the presence of a  presence of a PDR  around  this object class. While the physics
and chemistry of the low-excitation gas appears to be dominated by
photodissociation, a possible contribution due to shocks  can be inferred from the
evidence of gas phase Fe abundance enhancement.

\subsection{HD 168625}
Our mid-IR spectroscopic 
observations  (IRS) of this LBV candidate  detected spectral features attributable to  polycyclic aromatic hydrocarbons (PAHs), indicating 
the presence of a PDR around the ionized nebula.  This result
enlarges the number of LBV and LBV candidates where the presence of a PDR has been
confirmed, implying the importance of such a component in the budget of total mass lost
by the central object during this elusive phase of massive star evolution.

We have analyzed and compared the mid-IR and radio maps, and derive several results 
concerning the associated nebula (Umana et al. 2010).  While the overall torus-like shape of the dust morphology
is confirmed, the higher resolution and sensitivity of our images allow us to discern finer
details of the dust distribution, most notably the highly structured texture of the
nebula, and provide a better localization of the dust ring, with its north-west and south-east
condensations (Fig. 3).  There is also evidence for grain distribution variations 
across the nebula, with a predominant contribution from larger grains in the
northern part of the nebula while PAH and smaller grains are more segregated in the southern part.

Besides via optical emission, the  ionized part of the nebula can be  traced by radio observations, without suffering
of intrinsic extinction.
We have obtained a 3.6-$cm$  VLA map by using the interferometer in two configurations to 
determine the structure down to sub-arcsec scale without resolving out the more extended emission
(Fig 4). The overall ionized nebula is reminiscent of the dust distribution, with one main difference: 
 the brightest radio emission is located where there is a lack of thermal dust
grains,  corroborating the hypothesis of the presence of a shock in the southern portion of the nebula as
consequence of the interaction of a fast outflow with the slower, expanding dusty nebula.
Such a shock would be a viable means for PAH production as well as for changes in the grain
size distribution. Finally, from the detection of a central radio component, very probably associated with
the wind from the central massive supergiant, we derive a current mass-loss rate of 
$\dot {M}= (1.46 \pm 0.15) \times 10^{-6} M\odot$yr$^{-1}$. 

\subsection{Future prospects }
The study of the LBV phenomenon has been hampered by the lack of a significant sample of objects with associated nebulae.
The presence of an extended, dusty circumstellar nebula can be identified by its IR/mid-IR  fingerprints. Therefore,
we can  search  among proposed candidates by assessing the presence of observational  characteristics that define a LBV. 
A good possibility is offered by the more than 400 bubbles identified at $24\mu m$  by Mitzuno et al., 2010  in  the  Galactic Plane survey conducted with MIPS on the Spitzer Space telescope  (MIPSGAL). These small 
($\leq 1^{\prime}$ ) rings, bubbles, disks or shells are pervasive through the entire Galactic 
plane in the mid-infrared. Whatever the nature of these  $24\mu m$ sources, the implications of such 
a large number in the Galactic plane is remarkable and they provide a powerful  "game reserve" for evolved massive stars 
as already pointed out by Wachter et al (these proceedings).

\begin{figure}[h]
\begin{minipage}{8cm}
\centering
\includegraphics[width=9cm]{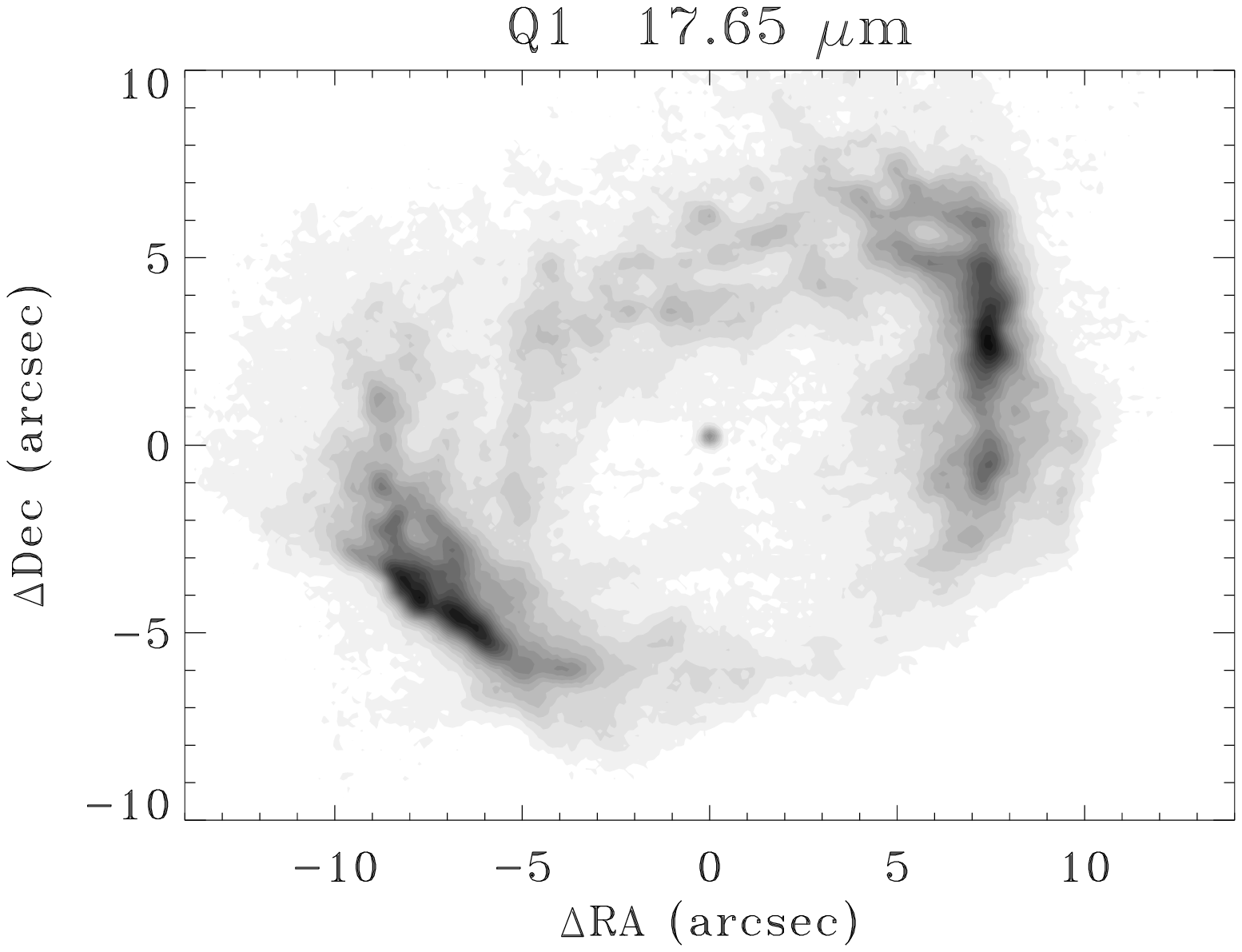}
\caption{VISIR map of HD~168625 in the Q1 continuum filter. 
The brightness levels range from 0 to 8.06 Jy~arcsec$^{-2}$. }
\end{minipage}
\hfill
\begin{minipage}{8cm}
\centering
\includegraphics[width=9cm]{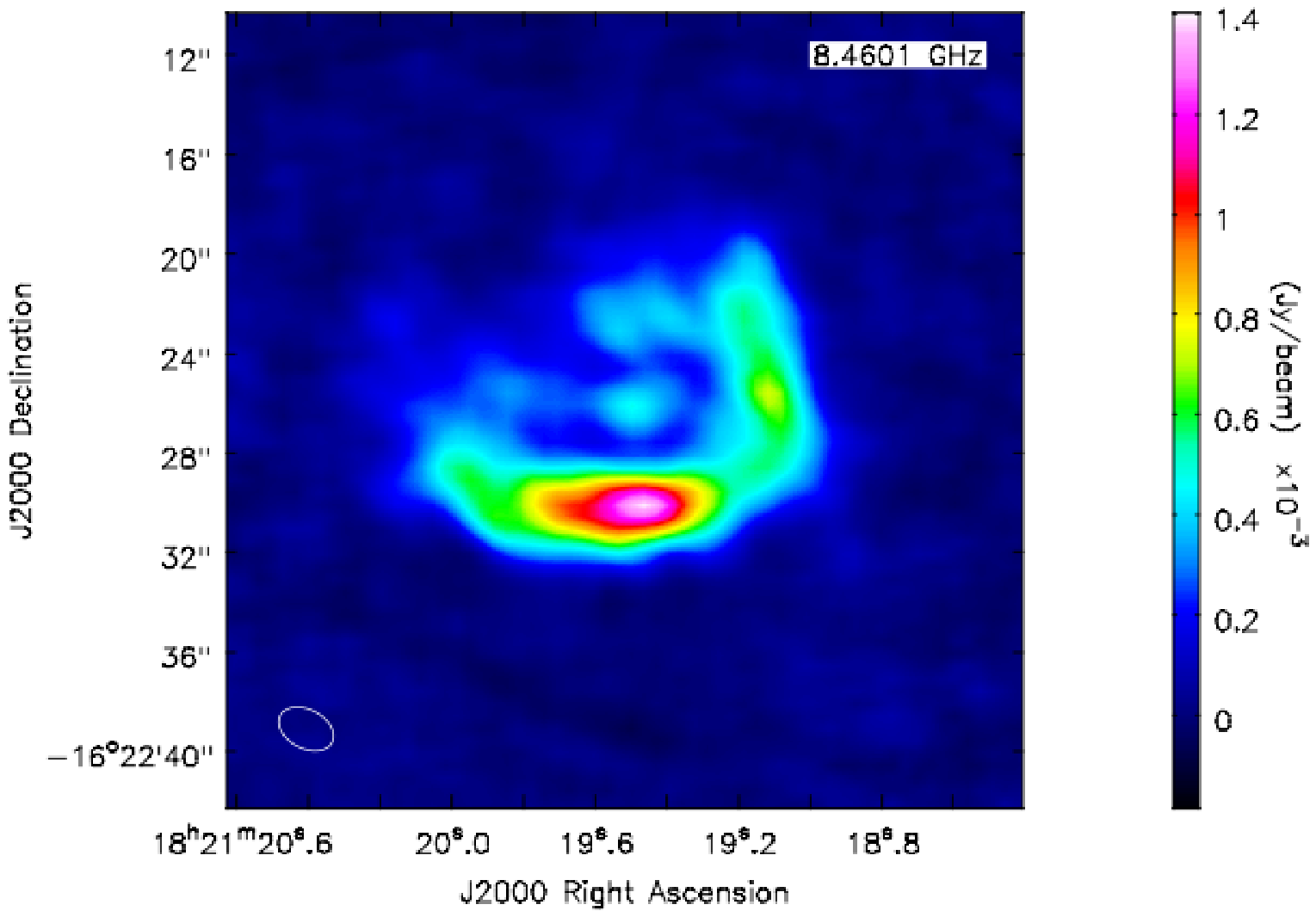}
\caption{The multi-configuration  3.6-$cm$  VLA map of HD168625.
The point-like central source, whose coordinates coincides with that of the central object, is probably related
to the current-day stellar wind of the LBV.}
\end{minipage}
\end{figure}

%
%
\section*{Acknowledgements}
This research is  supported in part by ASI contract
I/038/08/0 ``HI-GAL'' and by PRIN-INAF 2007. 
%
%
\footnotesize
\beginrefer
\refer Buemi, C. S., Umana, G.., Trigilio, C., Leto, P.,  Hora, J. L., 2010, ApJ, 721, 1404.

\refer Clark, J. S.,  Crowther, P. A.,  Larionov, V. M.,  Steele, I. A.,  Ritchie, B. W.,  Arkharov, A. A. 2009, A\&A 507, 1555

\refer Clark, J. S., Larionov, V. M.,   Arkharov, A., 2005, A\&A, 435, 239.

\refer Mizuno, D. R.,  Kraemer, K. E.,  Flagey, et al.,  2010, AJ 139, 1542.

\refer Umana, G.., Buemi, C. S.,  Trigilio, C.,  Leto, P.,  Hora, J. L., 2010, ApJ, 718, 1036

\refer Umana, G.., Buemi, C. S.,  Trigilio, C.,  Hora, J. L., Fazio, G. G., Leto, P.  2009, ApJ,  694, 697

\refer Umana, G.., Buemi, C. S.,  Trigilio, C.,  Leto, P.,  2005, A\&A , 437, L1.

\refer Waters, L. B. F. M., Morris, P. W.,  Voors, R. H. M.,  Lamers H. J. G. L. M. and Trams, N.R.  1998, AP\&SS, 255, 179.

\refer White, S. M., 2000, ApJ, 539, 851.

\endrefer           
\end{document}